\documentclass[sigconf, authorversion]{acmart}
\AtBeginDocument{%
  \providecommand\BibTeX{{%
    \normalfont B\kern-0.5em{\scshape i\kern-0.25em b}\kern-0.8em\TeX}}}

\copyrightyear{2023} 
\acmYear{2023} 
\setcopyright{acmlicensed}\acmConference[CHI '23]{Proceedings of the 2023 CHI Conference on Human Factors in Computing Systems}{April 23--28, 2023}{Hamburg, Germany}
\acmBooktitle{Proceedings of the 2023 CHI Conference on Human Factors in Computing Systems (CHI '23), April 23--28, 2023, Hamburg, Germany}
\acmPrice{15.00}
\acmDOI{10.1145/3544548.3580965}
\acmISBN{978-1-4503-9421-5/23/04}

\usepackage{xcolor}
\usepackage{xspace}
\usepackage{enumitem}
\usepackage{balance}

\newcommand{\eg}{\textit{e.g.}}
\newcommand{\ie}{\textit{i.e.}}
\newcommand{\etal}{\textit{et~al.}}

\def\tool{\textit{Notable}\xspace}

\newcommand{\revision}[1]{#1}

\newcommand{\haotian}[1]{#1}

\begin{document}

\title{\tool: On-the-fly Assistant for Data Storytelling in Computational Notebooks}
\author{Haotian Li}
\affiliation{%
  \institution{The Hong Kong University of Science and Technology}
  \city{Hong Kong SAR}
  \country{China}
}
\affiliation{%
  \institution{Microsoft Research Asia}
  \city{Beijing}
  \country{China}
}
\email{haotian.li@connect.ust.hk}

\author{Lu Ying}
\affiliation{%
  \institution{Zhejiang University}
  \city{Hangzhou}
  \state{Zhejiang}
  \country{China}
}
\affiliation{%
  \institution{Microsoft Research Asia}
  \city{Beijing}
  \country{China}
}
\email{yingluu@zju.edu.cn}

\author{Haidong Zhang}
\affiliation{%
  \institution{Microsoft Research Asia}
  \city{Beijing}
  \country{China}
}
\email{haizhang@microsoft.com}

\author{Yingcai Wu}
\affiliation{%
  \institution{Zhejiang University}
  \city{Hangzhou}
  \state{Zhejiang}
  \country{China}
}
\email{ycwu@zju.edu.cn}

\author{Huamin Qu}
\affiliation{%
  \institution{The Hong Kong University of Science and Technology}
  \city{Hong Kong SAR}
  \country{China}
}
\email{huamin@cse.ust.hk}

\author{Yun Wang}
\authornote{Yun Wang is the corresponding author.}
\affiliation{%
  \institution{Microsoft Research Asia}
  \city{Beijing}
  \country{China}
}
\email{wangyun@microsoft.com}


\begin{abstract}
Computational notebooks are widely used for data analysis. 
Their interleaved displays of code and execution results (\eg, visualizations) are welcomed since they
enable iterative analysis and preserve the exploration process.
However, the communication of data findings remains challenging in computational notebooks.
Users have to carefully identify useful findings from useless ones, document them with texts and visual embellishments, and then organize them in different tools.
Such workflow greatly increases their workload, according to our interviews with practitioners.
To address the challenge, we designed \tool to 
offer \textit{on-the-fly} assistance for data storytelling in computational notebooks.
It provides intelligent support to minimize the work of documenting and organizing data findings and diminishes the cost of switching
between data exploration and storytelling.
To evaluate \tool, we conducted a user study with 12 data workers.
The feedback from user study participants verifies its effectiveness and usability. 
\end{abstract}


\ccsdesc[500]{Human-centered computing~Visualization systems and tools}
\ccsdesc[500]{Human-centered computing~Interactive systems and tools}
\keywords{data visualization, data storytelling, computational notebooks}

\begin{teaserfigure}
  \includegraphics[width = \linewidth]{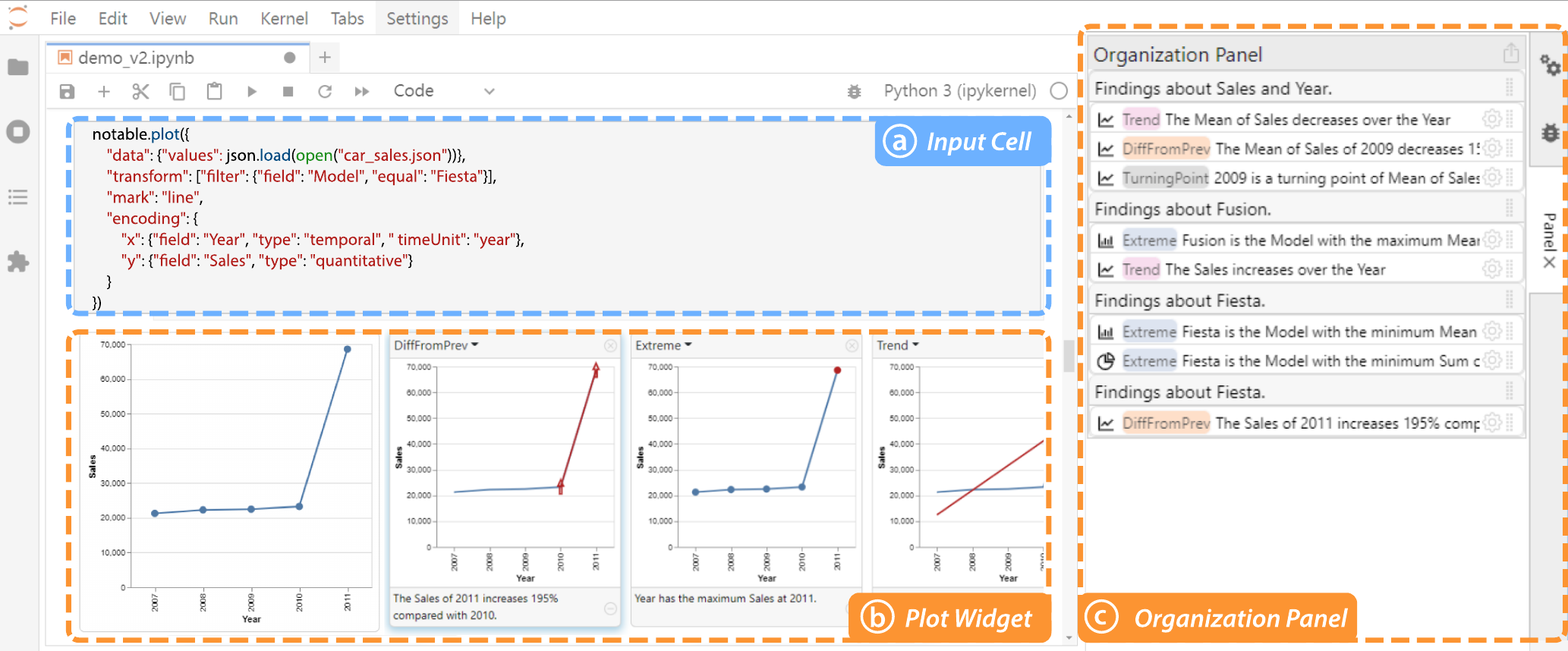}
    \caption{This figure illustrates the interface of \tool, an on-the-fly assistant for data storytelling in computational notebooks. (a) is an input cell where the chart is specified. The original chart and illustrated data facts are shown in (b), the \textit{plot widget}. The plot widget allows users to modify and select useful data facts.
    The selected data facts are arranged into a story and displayed in (c), the \textit{organization panel}.
    The automatically organized story can be further improved by users in the organization panel.}
    \label{fig:interface}
    \Description{This figure illustrates the interface of Notable, an on-the-fly assistant for data storytelling in computational notebooks. In the top left part of the figure, (a) is an input cell where the chart is specified. The original chart and illustrated data facts are shown in (b), the plot widget. (b) is in the bottom left part of the figure. The plot widget allows users to modify and select useful data facts. The selected data facts are arranged into a story and displayed in (c), the organization panel, in the right part of the figure. The automatically organized story can be further improved by users in the organization panel.}
\end{teaserfigure}

\maketitle

\section{Introduction}
Computational notebooks (e.g., Jupyter\footnote{\url{https://jupyter.org/}} and RStudio\footnote{\url{https://www.rstudio.com/}}) have been widely used in exploring data and deriving insights for decision making~\cite{rule2018exploration}.
Each computational notebook consists of multiple input and output \textit{cells} for code editing and result presentation~\cite{chattopadhyay2020painpoint}.
The design integrates the code and results in a single interface and thus suits the needs of data analysis: it allows iterative updates of code, facilitates quick inspection of the results~\cite{head2019managing}, and documents the procedure of data analysis for easier collaboration~\cite{rule2018exploration}.

Though computational notebooks have multiple advantages, they also introduce challenges to users
~\cite{chattopadhyay2020painpoint}.
Since users often explore data iteratively in computational notebooks, lots of intermediate or useless cells may be kept in the notebooks as well~\cite{head2019managing}.
Furthermore, the sequence of data exploration in computational notebooks may not entirely fit the narrative structure of a data story~\cite{kery2018story}.
Therefore, it is necessary for users to select the important findings, document the findings with text descriptions and visual highlights, and organize them into a logically coherent data story for communication~\cite{kery2018story}.
According to our interviews with experienced data analysts, such workflow often require them to use multiple tools (\eg, Microsoft PowerPoint\footnote{\url{https://www.microsoft.com/en-us/microsoft-365/powerpoint}} and Google Slides\footnote{\url{https://www.google.com/slides/about/}}), which increases their workload due to 
(1) \textit{the additional operations to transfer data findings between tools} and (2) \textit{the distraction led by multiple user interfaces with inconsistent appearances and interactions}.

To cater the need for convenient data communication, several existing tools are developed to support data storytelling.
For example, Viol\`{a}~\cite{viola} and Nbconvert~\cite{nbconvert} enable convenient format conversion from notebooks to presentations.
However, they only support direct format conversion and cannot facilitate other steps in data storytelling, such as story organization.
To fill the gap, a recent tool, NB2Slides~\cite{zheng2022telling}, adopted advanced machine learning techniques to organize an accomplished notebook into slides based on pre-defined templates.
However, it assumes that users finish the entire data analysis process before making data stories, which does not align well with the empirical observation of switching between data exploration and story creation~\cite{gratzl2016visual, lee2015more}.
Users can hardly consider data storytelling and exploration comprehensively in such settings.
Given the drawback of existing tools, we would like to explore \textit{how we can provide on-the-fly support to data storytelling during data exploration in computational notebooks.}

\haotian{To answer the research question, we first attempted to figure out the challenges and desired assistance regarding data storytelling in computational notebooks.
We conducted formative interviews with six data analysts with diverse backgrounds and summarized five design requirements of the storytelling tool in computational notebooks.
The interviewees frequently mentioned their need for assistance in data finding \textit{documentation and organization}.
Furthermore, they mentioned the \textit{inconvenient switching between data exploration and storytelling} when using computational notebooks.}
Based on their feedback, we designed and implemented \tool, a computational notebook extension, to offer on-the-fly assistance to storytelling, as shown in Figure~\ref{fig:interface}.
When the users plot a chart to inspect the data, \tool can automatically search for potential data facts that may interest the users and illustrate them with visual embellishments and text descriptions (Figures~\ref{fig:interface}(b)).
By doing so, the manual efforts of documenting findings can be eliminated.
Then the user can select a data fact, and \tool organizes the story based on the selected data facts and the new one (Figure~\ref{fig:interface}(c)).
The user may customize the data fact and the story organization as well.
At the end of data analysis, \tool supports exporting the story as presentation slides, 
one of the most commonly used formats of data stories~\cite{hullman2013deeper}.
To evaluate whether \tool is helpful to users, we conducted a user study with 12 participants.
The feedback from them demonstrates the usability and effectiveness of \tool.
Finally, we concluded our research by discussing the lessons learned and potential future directions.

To summarize, the contributions of our paper include:

\begin{itemize}
    \item The design requirements for storytelling tools in computational notebooks;
    \item A computational notebook extension, \tool, which offers on-the-fly assistance to data storytelling;
    \item The design lessons and future opportunities learned from the research.
\end{itemize}
\section{Related Work}
In this section, we review research on visual analysis in computational notebooks, data storytelling, and data fact recommendation.

\subsection{Visual Data Analysis with Computational Notebooks}
Computational notebooks have been widely applied in data analysis since it enables the integrated display of code and results, which facilitates the iterative nature of data analysis better~\cite{rule2018exploration}.
In existing computational notebook environments, visual data analysis is supported by packages for convenient plotting (\eg, Altair~\cite{vanderplas2018altair}, Bqplot~\cite{bqplot}, Plotly~\cite{plotly}, and Matplotlib~\cite{hunter2007matplotlib}) and \revision{tools targeting at visualizing specific machine learning models (\eg, TimberTrek~\cite{wang2022timbertrek} and Calibrate~\cite{xenopoulos2022calibrate})}.

The aforementioned plotting packages are widely used but limited to showing the visualizations specified by the user and do not provide additional assistance.
Recent studies have been proposed to augment them by providing intelligent assistance to data exploration~(\eg, \cite{lee2021lux, zhang2019mining, chen2022pi2, wang2022diff, wu2020b2}).
For example, Lux~\cite{lee2021lux} recommends lists of static visualizations to users without requiring them to specify visualizations explicitly.
Considering the importance of interactions in visual analysis~\cite{yi2007toward}, PI$_1$ recommends interactive visualizations when users query data from databases~\cite{zhang2019mining}.
Informed by the advantages of multi-view visualizations over single-view charts~\cite{lin2022dminer}, PI$_2$~\cite{chen2022pi2} extends PI$_1$ by generating multi-view interactive visualizations.
Besides recommending visualizations, EDAssistant~\cite{li2021edassistant} and LodeStar~\cite{raghunandan2022lodestar} suggest code snippets for data exploration.
\revision{Fork It~\cite{weinman2021fork} allows users to keep multiple versions of code in notebooks for convenient alternative explorations.}

Though these tools assist users in data exploration by recommendation, they still require a considerable amount of effort to communicate the findings, which is an essential stage in data analysis~\cite{wongsuphasawat2019goals}.
Existing extensions allow users to directly convert notebooks into other formats for presentation, such as Voilà~\cite{viola} and Nbconvert~\cite{nbconvert}.
\haotian{However, they only help format conversion and cannot cover other steps in data storytelling.}
In our research, we propose \tool, a notebook extension, to facilitate data storytelling during data exploration.
Despite generating slides for presenting the data story, \tool also offers assistance in (1) illustrating data findings with text description and visual embellishments; and (2) organizing selected data facts into stories.

\subsection{Data Storytelling}
Data storytelling, as an effective way to communicate information in data, has gained more attention in recent years~\cite{tong2018storytelling}.
It serves as the final stage in data analysis and is closely connected with data exploration~\cite{wongsuphasawat2019goals}.
To facilitate data storytelling, various types of tools have been proposed.
According to the level of automation, they can be roughly classified into three clusters, authoring tools, automatic generation tools, and tools with intelligent support~\cite{chen2022survey}.

Authoring tools provide interactive interfaces for users to create data stories freely.
For example, Idyll Studio~\cite{conlen2021idyll} and VizFlow~\cite{sultanum2021leveraging} support data article creation by integrating chart creation and article editing in a unified interface.
CLUE~\cite{gratzl2016visual} and InsideInsights~\cite{mathisen2019insideinsights} allow users to create data presentations with the identified findings in data exploration.
ToonNote~\cite{kang2021toonnote} integrates a data comic authoring interface with computational notebooks.
To reduce the considerable manual efforts in creating data stories, fully automatic data story generation tools have been explored.
They leverage machine learning techniques to analyze datasets and generate narrative visualizations.
Under this category, 
Wang~\etal~\cite{wang2019datashot}, Shi~\etal~\cite{shi2021autoclips}, and Lu~\etal~\cite{lu2021automatic} automate fact sheet, data video and scrollytelling creation from datasets.
\revision{Similarly, Roslingifier~\cite{shin2022roslingifier} detects important events in time series data and generates animation automatically.}
Chen~\etal~\cite{chen2020supporting} attempted to synthesize data stories according to the analytical provenance.
InfoMotion~\cite{wang2021animated} creates animated infographics according to the structure of information inside infographics.
However, automatic story generation tools limit users' participation in storytelling and thus may result in bias and untrust~\cite{li2021exploring}.

To achieve the collaboration of humans and machines, tools that provide intelligent support to data storytelling have gained increasing interest ~(\eg, \cite{yuan2021infocolorizer, obie2022gravity++, ge2021cast, winters2019automatically, sun2022erato}).
\revision{For example, Erato~\cite{sun2022erato} recommends new data facts based on user-selected facts for completing data stories and generating infographics.}
According to a recent survey~\cite{chen2022survey}, most of the existing tools support infographic or video and animation creation, while making slides receives little attention.
According to Hullman~\etal~\cite{hullman2013deeper}, slides are frequently used for communicating data stories.
However, making slides requires non-trivial efforts from data scientists~\cite{piorkowski2021ai}.
To assist with slide creation, Zheng~\etal~\cite{zheng2022telling} developed NB2Slides to convert human-made computational notebooks to slides based on pre-defined templates.
However, NB2Slides takes a one-way style that generates slides after a complete data analysis session.
Such features may not suit the common workflow of data analysts as they often switch between data story creation and data exploration, according to previous research~\cite{gratzl2016visual, lee2015more}.
To better fit the workflow, we propose to offer on-the-fly assistance in \tool to support story creation during data exploration.
Users are able to take storytelling into consideration when exploring data.
\revision{Furthermore, our tool is integrated into widely used computational notebooks, which is preferred over new tools, such as Erato~\cite{sun2022erato}, as indicated in our formative study.}

\subsection{Data Fact Recommendation}\label{sec:data_fact_rec}
Since manual visual data exploration requires data analysis skills and enormous effort~\cite{alspaugh2018futzing}, recent research introduces automatic \textit{data fact} recommendation methods to address this challenge~\cite{law2020characterizing}.
These methods examine data characteristics automatically and recommend data facts that may interest users, \eg, outliers and trends of data.
Data facts are also called \textit{data insights} in other publications~(\eg, \cite{tang2017extracting, ding2019quickinsights}).

The purpose of recommending data facts can be diverse~\cite{law2020characterizing}.
For example, QuickInsights~\cite{ding2019quickinsights}, Top-K Insights~\cite{tang2017extracting}, and SeeDB~\cite{vartak2014seedb} recommend data facts to facilitate a quick exploration of a database before in-depth data analysis.
Duet~\cite{law2018duet} suggests similar data facts when analyzing a specific subset of data, which happens in the data analysis with a clear target.
The line of research that inspires our study applies data facts in data communication (\eg, \cite{wang2019datashot, shi2020calliope}).

\haotian{Regardless of the purposes, most  existing studies above attempt to explore data column combinations and recommend potentially interesting data facts from a dataset.}
\haotian{However, most data workers still prefer to plot charts and observe data manually instead of relying on the recommendation~\cite{alspaugh2018futzing}.
To facilitate their needs, in this paper, we explore how to reduce the workload of data exploration based on the charts created by them.}
In our formative interviews, we identified that manually documenting data findings in a chart required considerable effort.
Users need to write text descriptions and sometimes highlight key data points.
To automate this procedure, \haotian{we propose to use the algorithm of data fact recommendation to infer the potential data facts from the charts that users have created and may feel interested in.} 
Then we illustrate the data facts for users' selection to facilitate data insight interpretation and communication \cite{srinivasan2018augmenting}.
According to users' feedback, the illustrated data facts can save their time and facilitate storytelling.

\section{Formative Study}\label{sec:formative}
To derive the design requirements of storytelling tools in notebooks, we conducted a formative study where we interviewed data workers from different domains in a semi-structured manner.

\subsection{Interviewees}
In our formative study, we recruited six data workers (3 male and 3 female, $Age_{mean} = 27.5$, $Age_{std} = 3.5$, $Experience_{mean} = 4.67~years$ , $Experience_{std} = 0.47~year$) by sending invitations through social media.
They were from both academia and industry with diverse backgrounds.
They had at least 4-year experience in using computational notebooks for data analysis.
All of them explored data and communicated data findings frequently (\ie, at least once a week).
Their demographic information is in Table~\ref{tab:demographic}.

\begin{table*}[h!]
    \centering
    \caption{The table records the demographic information of the interviewees in our formative study. We report their genders, ages, jobs, domains, experiences in years, and frequency of data exploration and communication. }
    \begin{tabular}{@{}lllllll@{}}
    \toprule
    ID & Gender & Age & Job & Domain & Experience \revision{(Year)}& Frequency\\ \midrule
      P1 & Female & 26 & Postgraduate researcher & Data visualization & 4 & Everyday \\
      P2 & Male & 28 & Data analyst & Finance & 5 & Twice a week \\
      P3 & Female & 26 & Research fellow & Computational social science & 4 & Everyday\\
      P4 & Male & 25 & Postgraduate researcher & Economics & 5 & Everyday\\
      P5 & Male & 35 & Applied data scientist & Computer system & 5 & Everyday\\
      P6 & Female & 25 & Data scientist & E-commerce & 5 & Once a week
       \\\bottomrule
    \end{tabular}
    
    \label{tab:demographic}
\end{table*}

\subsection{Procedure}
Our study was conducted through one-on-one online meetings.
Each meeting began with an introduction to computational notebooks to recall interviewees' experiences.
Then we asked for the interviewees' consent to recording the meeting and using their demographic information before the interview.
The semi-structured interview study had three parts.
In the first part, the nature of the interviewees' data work was enquired to learn their background.
For example, we asked them \textit{``what is your daily work?''}
In the second part, 
interviewees introduced their workflow of communicating data exploration results in computational notebooks.
We also asked about their pain points in the current workflow.
In the last part, the interviewees discussed how to improve their workflow with us.
Each study lasted about half to an hour. The authors took notes during the meetings.
After all interviews, the first author organized the feedback according to the notes and video recordings.
Then the co-authors discussed and summarized the organized feedback into findings iteratively with the help of recordings.
Finally, five design requirements were derived from the findings.
\subsection{Findings}
All participants agreed that it took considerable effort to make data stories based on findings in exploration.
According to our interviewees, when data findings were observed, 
they first \textit{documented} useful ones
with texts or visual embellishments.
Then the documented findings would be \textit{organized} into data stories.
To present data stories, the interviewees needed to \textit{prepare} slides.
We summarize three key findings regarding the pain points and their expected tools. 

\subsubsection{Manual documenting and organizing data findings}
In the interview, all interviewees mentioned the difficulties of \textbf{documenting the discovered data findings} during data exploration.
First, they had to manually record the data findings and highlight the key data points on the chart.
P5 said that he would like to highlight the key findings on the charts but often found that it was challenging to do it by programming.
Second, five among six interviewees had to rely on additional tools to document their data findings. 
P2 took notes using paper, while others used note-taking applications such as Notion\footnote{\url{https://www.notion.so/}} or OneNote\footnote{\url{https://www.microsoft.com/en-us/microsoft-365/onenote/digital-note-taking-app}}.
Only P3 used the markdown cells in computational notebooks to record the findings.
Due to the two problems, documenting data findings when using notebooks often distracted them from data exploration, which has been mentioned in a previous study~\cite{wang2022documentation} as well.

Despite the documentation of data findings, \textbf{the organization of findings} was also mentioned by four interviewees (P1-P3, P5) as a pain point.
P1 commented that she documented a large number of data findings during the exploration.
It was hard for her to remember the relationship between the findings. 
Therefore, the story organization was highly challenging, which echoed the observation in previous research~\cite{kery2018story}.
She hoped that the data findings could be automatically organized according to their common features.
P3 complained that she needed to \textit{``spend half of the time in data exploration and another half in preparing presentations''}, indicating the great effort she took to organize data findings and make slides.

\subsubsection{Inconvenient switching between data exploration and storytelling}
All six participants commented that they commonly switched between data exploration and storytelling.
P5 mentioned \textit{``organizing data findings is like organizing my mind''}.
He could notice logical flaws in his data analysis during authoring data stories and then went back to data exploration.
P4 also said \textit{``I repeatedly iterate between exploration and storytelling''} since he could get new ideas when preparing his stories.
The feedback reveals that storytelling is bi-directionally connected with data exploration, which aligns with previous observations~\cite{lee2015more, gratzl2016visual, wongsuphasawat2019goals}.
Making the story also inspires and guides the process of data exploration.

Five interviewees~(P1-P4, P6) complained about the inconvenience of switching between data exploration and storytelling when using computational notebooks.
The issue was led by \textbf{the usage of multiple tools}.
When analyzing data, they explored datasets with computational notebooks and identified useful findings.
These findings in exploration were documented using various approaches, such as writing on paper or using note-taking applications.
They commonly organized findings and prepared slides using presentation tools, including Microsoft PowerPoint and Google Slides.
The interviewees indicated two-fold drawbacks led by the issue.
First, it is not easy to transfer data findings between tools.
Three interviewees (P1, P2, P5) mentioned that they had to take screenshots of charts and import them to the presentation tool.
P6 even needed to copy and paste data between tools manually.
The comments aligned with a previous study~\cite{brehmer2021jam}, where the authors describe transferring data findings as a \textit{``tedious''} process.
Second, the usage of different tools introduced additional mental load and led to distraction.
For example, P1 felt that her exploration was interrupted when she needed to transfer the findings from computational notebooks to another application.

\subsubsection{Design of storytelling tools}
At the end of the interviews, the participants were encouraged to describe their expectations about future storytelling tools.
All participants welcomed intelligent assistance to address their challenges.
Besides, they have two common suggestions.
First, four interviewees (P1, P2, P5, P6) emphasized \textbf{the necessity of convenient customization} in storytelling tools.
P2 worried that existing techniques for automatic data exploration and story generation were only able to mine the apparent data patterns. His team usually looked for the causes of data patterns leveraging domain expertise, and therefore it was important to let users involve in story content creation.
P1 pointed out that different users could have different approaches to organizing the findings into stories.
P6 further mentioned the necessity of exporting modifiable formats of stories since most companies have requirements on the appearance of stories.
Therefore, it is crucial to enable users' customization, and the customized story should have higher priority than the automatically organized story.
Second, the tools are better to be \textbf{integrated with existing tools}, \ie, existing computational notebook environments and commonly used presentation tools.
P1 expressed her doubts about an entirely new tool due to the learning curve.
She preferred an extension with simple interactions so that she would not spend extra effort on learning the usage of the tool.
P4 echoed P1's opinion by saying \textit{``it will be great if the tool is an extension and can be integrated into widely accepted tools''}.

\subsection{Design requirements}
We derived five key design requirements of a storytelling tool in the computational notebook according to the findings. In the remaining parts of the paper, \textbf{R1}-\textbf{R5} refer to the requirements. The requirements are:

\textbf{R1. Offering on-the-fly assistance to data storytelling.}  
The storytelling tool should support concurrent data storytelling and exploration to eliminate the cost of switching. With the tool, users can swiftly transfer data findings into story pieces; users can also reflect on their exploration with data stories.

\textbf{R2. Facilitating data finding documentation.}
The tool should assist the user in finding documentation.
It should reduce the work of illustrating findings with text descriptions and visual embellishments.
Furthermore, the illustrated findings should be easily accessed in the tool.

\textbf{R3. Automating the organization of documented data findings.}
The organization of documented data findings should be automated to reduce the users' workload.
The organized story can facilitate the quick reflection of data exploration.
Users can identify potential logical flaws and unexplored data subsets through the organized story.

\textbf{R4. Supporting customization of data stories.} 
Users should be allowed to customize their data stories conveniently.
The tool should allow users to revise story content and organize stories freely in the tool.

\textbf{R5. Integrating with existing tools.} 
The storytelling tool should be integrated with common computational notebook environments and presentation tools.
Therefore, users do not need to adjust their formed habits.
They can explore data and improve the data stories with familiar computational notebooks and presentation tools.

\begin{figure*}[h!]
    \centering
    \includegraphics[width = \linewidth]{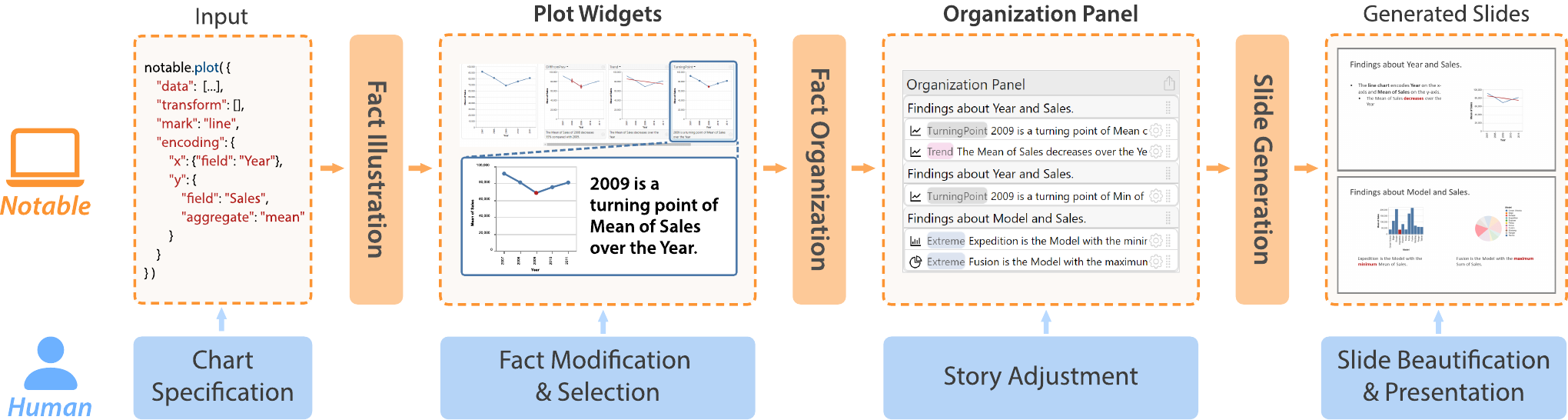}
    \caption{This figure shows the workflow of \tool. \tool contains five major modules: fact illustration, fact organization, slide generation, plot widget, and organization panel. The last two are displayed in the computational notebook interface.}
    \label{fig:overview}
    \Description{This figure shows the workflow of Notable. The workflow has seven stages from left to right, input, fact illustration, plot widgets showing illustrated facts, fact organization, organization panel showing organized facts, slide generation, and generated slides. Users have four actions in the workflow, chart specification, fact modification and selection, story adjustment, and slide beautification and presentation.}
\end{figure*}

\section{\tool}
In this section, we first present an overview of \tool(Section~\ref{sec:overview}). Then we introduce the definition of data facts~(Section~\ref{sec:data_fact}) and the modules that support data storytelling during data exploration~(Sections~\ref{sec:algorithms} and~\ref{sec:interface}).

\subsection{Overview}\label{sec:overview}
\tool is a computational notebook extension that provides assistance to data story authoring during data exploration.
Following \textbf{R5}, it is integrated with one of the most widely used computational notebook environments, JupyterLab.
The tool supports exporting data stories as PowerPoint files, which facilitates follow-up modification and presentation.

Figure~\ref{fig:interface} shows how \tool looks in the notebook interface.
It composes of multiple \textit{plot widgets}~(\eg, Figures~\ref{fig:interface}(b)) and an \textit{organization panel}~(Figure~\ref{fig:interface}(c)) for each notebook.
Each plot widget presents the user-created chart and illustrated data facts. 
The organization panel presents the organized data facts and allows the user to adjust the story organization.
Plot widgets and the organization panel are the \textit{interactive modules} in \tool.
These modules, together with the input cells in computational notebooks, enable exploring datasets and creating data stories in a unified tool~(\textbf{R1}).

To support the functionalities of interactive modules, three \textit{computation modules}, \textit{fact illustration}, \textit{fact organization}, and \textit{slide generation}, are designed.
Their relationship is illustrated in Figure~\ref{fig:overview}.
The fact illustration module accepts users' chart specifications in Vega-Lite~\cite{satyanarayan2016vegalite} as the input and illustrates the potential data facts.
The advantage of Vega-Lite is that it is a high-level visualization specification grammar with a relatively low learning cost.
In the future, it is possible to support other approaches of chart specification, \eg, Matplotlib~\cite{hunter2007matplotlib}.
The illustrated facts are displayed in a plot widget.
Once a data fact is selected by the user, the fact organization module suggests a potential position of the fact in the data story.
The organization panel displays the current data story that the user is working on.
After the story is compiled, the slide generation module generates a slide deck for further editing.
The three modules enable transferring data facts from exploration to storytelling seamlessly~(\textbf{R1}).

\subsection{Data Fact Definition}\label{sec:data_fact}
In \tool, facts are considered as the basic units in a data story. \tool first identifies the facts in specified charts and illustrates them.
Then the selected facts are organized into a story.
Following a previous study~\cite{wang2019datashot}, we characterize data facts using seven attributes: 
\revision{\textit{subspace} to record the filters that are applied to gain the visualized data subset; \textit{measures} to indicate the dependent variables in the chart; \textit{dimension} to represent the independent variable in a chart; \textit{type} of data facts (\eg, trend and outlier); \textit{parameters} to describe the details of the fact, such as the direction of trends; \textit{focus} to document the data point that is emphasized in the fact; and \textit{score} to measure both how the fact matched users' intent of exploration and how important the fact is}.

\subsection{Computation Modules}\label{sec:algorithms}
In this section, we introduce the computation modules that support the interactive modules.

\subsubsection{Fact Illustration}\label{sec:fact_illustration}
According to our interviews and previous research~\cite{wang2022documentation}, documenting data findings is one of the pain points in storytelling with computational notebooks.
To tackle the issue, we propose to illustrate facts automatically to reduce the workload of manual documentation (\textbf{R2}).
If users are interested in some illustrated facts, they can directly add them to stories without additional operations.
The added facts will be shown in the organization panel for convenient inspection later.
To achieve automatic fact illustration, it is essential to infer what fact types the user is interested in and then identify those important facts.
We adopt fact mining algorithms~\cite{ding2019quickinsights, wang2019datashot} to extract facts that will be illustrated. 

When a chart is plotted using \tool, the subspace, the measure, and the dimension are extracted directly from its specification.
With the three attributes, \tool first transforms the input dataset by applying filters and aggregations.
Then \tool attempts to search for potential data facts in the transformed dataset. 
For example, it identifies whether outliers exist using the widely recognized three-sigma rule\footnote{\url{https://encyclopediaofmath.org/wiki/Three-sigma_rule}} and conducts regression to check the existence of trends.
After this step, \tool constructs a collection of potential data facts with their fact types, focuses, and parameters.
At the end of fact computation, all potential facts are sorted according to their scores, and top-\textit{k} facts among all are illustrated to users.
\textit{k} is set to 3 by default due to the limited screen space and can be configured by users.
\revision{To ensure the diversity of recommended data facts, \tool first selects the data facts with the highest score of every type and recommends the top-$k$ facts.
If the number of recommended facts is less than $k$, the other facts will be sorted in descending order according to scores and be recommended until $k$ facts are presented.}

In previous research, the score of a fact represents its importance and is composed of two parts, \ie, \textit{impact} and \textit{significance}.
\textit{Impact} represents the coverage of the data subspace over the entire dataset, while \textit{significance} measures how obvious the data fact is.
For example, DataShot~\cite{wang2019datashot} computes the score of a fact as $score = \sum_{i \in [significance, impact_f, impact_c]}w_i * score_i$, where $impact_f$ is the impact of the focus and $impact_c$ is the impact of the context.

However, the previous definition of fact score does not entirely fit our scenario.
We would like to infer users' intent of exploration when plotting charts.
Therefore, we further consider the \textit{suitability} score between the fact type and the chart type.
The selection of chart type can be related to the users' analysis and presentation purposes since different chart types are effective for different purposes~\cite{saket2018task}.
For example, when a line chart is plotted, the creator may care more about the trend of data instead of the fact where a data point occupies the majority of the overall values.
The suitability score is computed as the probability of representing a fact using a certain chart type.
The probability is derived from the statistics between fact types and chart types by Wang~\etal~\cite{wang2019datashot}.
Their statistics summarize the usage of chart types against fact types in data stories.
\revision{For example, according to the statistics, 42 in 57 trend facts are represented with line charts. 
Then the suitability score of illustrating a trend fact in a line chart is $42/57=0.74$.}
Since the ultimate goal of our paper is data storytelling, it is more suitable to use the relationship between fact types and chart types in data stories rather than in other scenarios (\eg, \cite{saket2018task}).

\begin{table*}[]
\caption{The table presents templates of text descriptions generated by the fact illustration module.}
\begin{tabular}{ll}
\toprule
Fact Type & Template \\ \midrule
     
        Majority  &     The category of \{focus\} accounts for the significant amount \{ratio\} of \{measure\}.             \\
       Extreme   &     \{dimension\} has the maximum/minimum \{measure\} at \{focus\}.           \\ 
       Outlier  &     \{dimension\} has an outstanding \{measure\} at \{focus\}.            \\
       Turning point & \{focus\} is a turning point of \{measure\} over the \{dimension\}. \\
       Difference     &     The \{measure\} of \{$\textrm{focus}_\textrm{1}$\} increases/decreases \{ratio\}\ compared with \{$\textrm{focus}_\textrm{2}$\}.     \\
       Trend & The \{measure\} increases/decreases over the \{dimension\}. \\
       \bottomrule
\end{tabular}
\label{tab:text_template}
\end{table*}

Furthermore, the impact of context is meaningless in our setting since all facts derived from one chart have the same context.
\revision{Therefore, we only keep the impact of focus and calculate it as the proportion of data points in the focus over the dataset.
For example, if the focus is a turning point in a dataset of five rows, its $impact_f$ is $1/5 = 0.2$.
Since significance scores concern data patterns, we mainly follow QuickInsights~\cite{ding2019quickinsights} and DataShot~\cite{wang2019datashot} to compute them and the computation methods depend on fact types.
For example, the significance score of difference facts is the normalized relative difference between two data points.
In a data column, $(1, 5, 15, 14)$, the relative differences between consequent data points are 4, 2, and 0.07.
Therefore, $significance$ of the difference fact regarding $(5, 15)$ is $(2-0.07)/(4-0.07) = 0.49$.}
To consider three scores jointly, \tool computes the fact score as $score = \sum_{i \in [significance, impact_f, suitability]}w_i * score_i$.
\revision{Based on the previous approach of parameter selection~\cite{wang2019datashot}, we empirically set the weights to be 0.5, 0.2, and 0.3, respectively.
The weights are adjustable to fit personal preferences.
For example, if a user prefers not to consider the suitability of charts, $w_{suitability}$ can be set to 0.
}

\begin{figure}[h!]
    \centering
    \includegraphics[width=\linewidth]{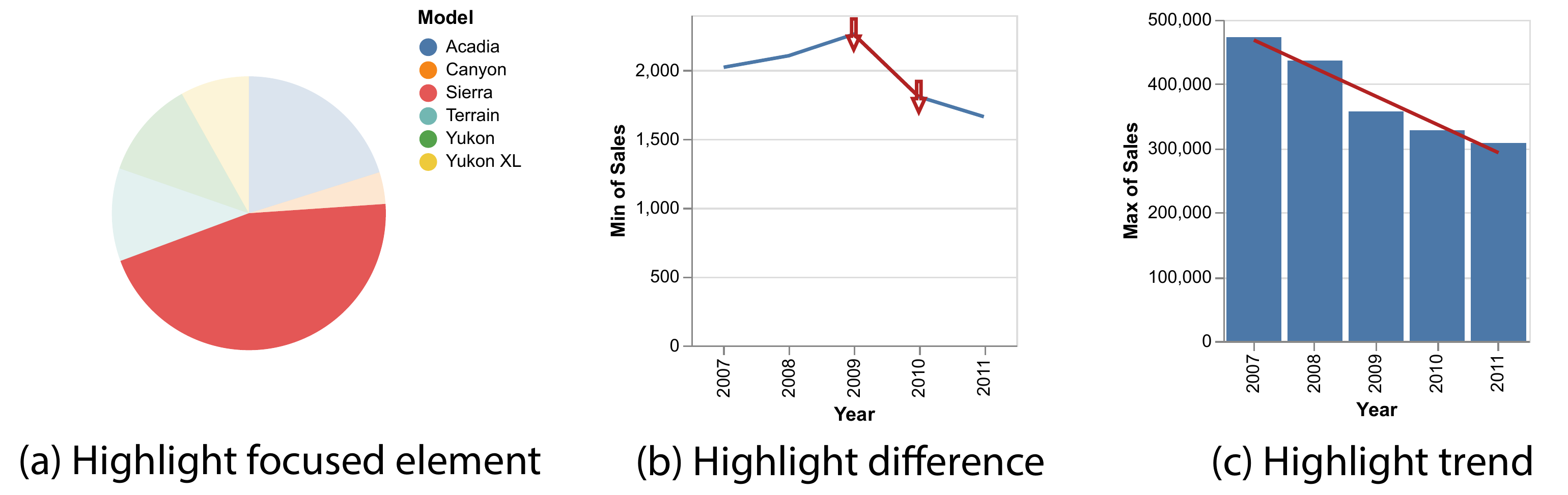}
    \caption{This figure presents how \tool illustrates data facts with different visual embellishments.}
    \label{fig:highlight}
    \Description{This figure presents how Notable illustrates data facts with different visual embellishments. It highlights one focused element in (a) on the left or two elements to show the difference between them in (b) in the middle. It also highlights a line for showing a trend in (c) on the right.}
\end{figure}

Based on extracted facts, we illustrate the original chart with text descriptions and visual embellishments.
The usage of descriptions and embellishments is intended to  reduce users' effort in documenting data facts and facilitate understanding data facts~\cite{srinivasan2018augmenting}.
Following previous practice in generating explanations for visualization~\cite{liu2020autocaption}, we design templates for different types of data facts.
The list of templates are shown in Table~\ref{tab:text_template}.
These templates contain information about dimension, measure, focus, type, and parameters.
The reason why subspace is not included by default is that we would like to keep the text description simple and concise, which aligns with slide design rules~\cite{green2021basics}.
Users can enable subspace in the description as well.
Furthermore, we design three types of visual embellishments for different types of data facts.
First, \tool highlights the focused data point in data facts that only consider a single value, such as extreme and outlier (Figure~\ref{fig:highlight}(a)).
Second, to handle facts showing the difference between two data points, \tool links two data points and highlights both data points with two arrows that indicate the direction of differences, \ie, increasing and decreasing (Figure~\ref{fig:highlight}(b)).
At last, since the fact regarding data trends considers all data points, \tool adds an additional trend line to demonstrate the trend  (Figure~\ref{fig:highlight}(c)).
The visual highlights are added to the charts by modifying the chart specification, which facilitates potential manual chart improvement.
Then the illustrated data facts, together with the original chart, are passed to the plot widgets for users' inspection.

\subsubsection{Fact Organization}\label{sec:fact_organization}
\haotian{Users are allowed to freely organize their identified data facts in \tool.
However, according to our interviewees, organizing facts into data stories requires considerable effort.
To reduce the effort, we provide the fact organization module to suggest a potential arrangement of facts according to the data relationship~(\textbf{R3}).}

After an illustrated data fact is selected through plot widgets, \tool organizes all selected data facts at once.
\tool first goes through the entire slide deck and searches for a suitable slide where the fact can be inserted.
In our algorithm, a suitable slide has two criteria: (1) all facts in the slide are identified in the same chart; (2) the slide has less than three facts.
The rationale behind the two criteria is to minimize the diversity and quantity of information in one slide~\cite{green2021basics}.
If a suitable slide exists, the new fact will be inserted, and the sequence of facts in the slide will be re-arranged.
If there is no suitable slide, a new slide will be created for the fact, and the sequence of slides will update.

\begin{figure*}[!h]
    \centering
    \includegraphics[width=0.75\linewidth]{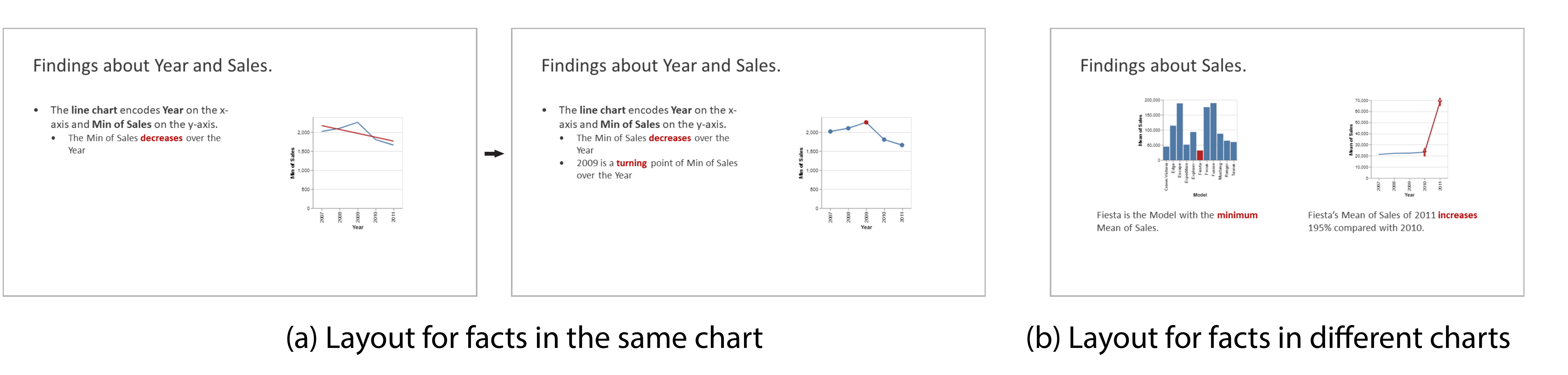}
    \caption{This figure shows two layout designs in the slide generation module.}
    \label{fig:layout}
    \Description{This figure shows two layout designs in the slide generation module. (a) on the left shows the layout for facts in the same chart. (b) on the right shows the layout for facts in different charts.}
\end{figure*}

To arrange the sequence of facts, we follow Hullman~\etal~\cite{hullman2013deeper} and Kim~\etal~\cite{kim2017graphscape} to minimize the sum of transition costs between adjacent facts in the fact sequence.
Each transition cost is calculated based on several factors mentioned in previous research~\cite{hullman2013deeper}, including the temporal relationship between facts and the consistency between fact focuses.
Similarly, \tool minimizes the transition costs between slides.
When computing the transition costs between slides, one issue is that a slide may contain facts based on different data attributes or subspaces. 
The issue appears when the user manually places facts based on different charts in one slide.
In such a situation, \tool cannot directly estimate the transition costs led by changes in data attributes between slides.
Inspired by DataShot~\cite{wang2019datashot}, \tool first extracts the shared measure, dimension, subspace, and focus as the \textit{topic} of the slide.
The topic of a slide can reveal the core idea of the slide.
Therefore, we are able to estimate the transition cost between two slides as the transition costs between topics using the approach in previous research~\cite{hullman2013deeper}.
The slide title is also generated with the shared data attributes in its topic.
Another special consideration is the sequence of chart creation in the notebook.
The chart sequence may reveal the user's logic flow.
To preserve users' logic flow in the organized story, \tool considers chart position relationship in computing transition cost as well.

Since the users are allowed to adjust the sequence of facts and slides manually, the fact organization module follows the principle that the users' actions prioritize automatic organization~(\textbf{R4}).
Once the sequence has been updated by users manually, the newly inserted slides or facts will not affect the manually arranged sequence.

\subsubsection{Slide Generation}
The last computation module generates slides when users feel comfortable with the organized story.
The slide generation module can export the story as PowerPoint slides.
By doing so, we aim to allow users to further edit slides with familiar tools, \eg, changing the slide template and adding animation~(\textbf{R4, R5}).
Users can also easily re-use the results of data exploration without opening notebooks again.

The generated slides have two layouts to accommodate different arrangements of facts.
The first layout is designed to present multiple facts observed in the same chart progressively (Figure~\ref{fig:layout}(a)).
On the left, an introduction to the encoding of charts is placed at the top.
Then the facts are introduced one by one.
The chart on the right changes as a new description of the fact is added.
The design aims to let the audience focus on one fact at one time.
The second layout presents facts derived from different charts (Figure~\ref{fig:layout}(b)).
The side-by-side design can better facilitate the need to present the relationship between two facts.
Due to the limited screen space, the second layout does not have an introduction to chart encodings.
The slide generation module also highlights key texts in facts such as the fact types and fact parameters.
In this way, the data facts can be better conveyed to the audience~\cite{green2021basics}.

\subsection{Interactive Modules}\label{sec:interface}
This section introduces the interactive modules in the notebook interface.
They present the results of the computation modules to users and enable seamless story creation~(\textbf{R1}) and customization during data exploration in computational notebooks~(\textbf{R4}).

\begin{figure*}[h]
    \centering
    \includegraphics[width=0.75\linewidth]{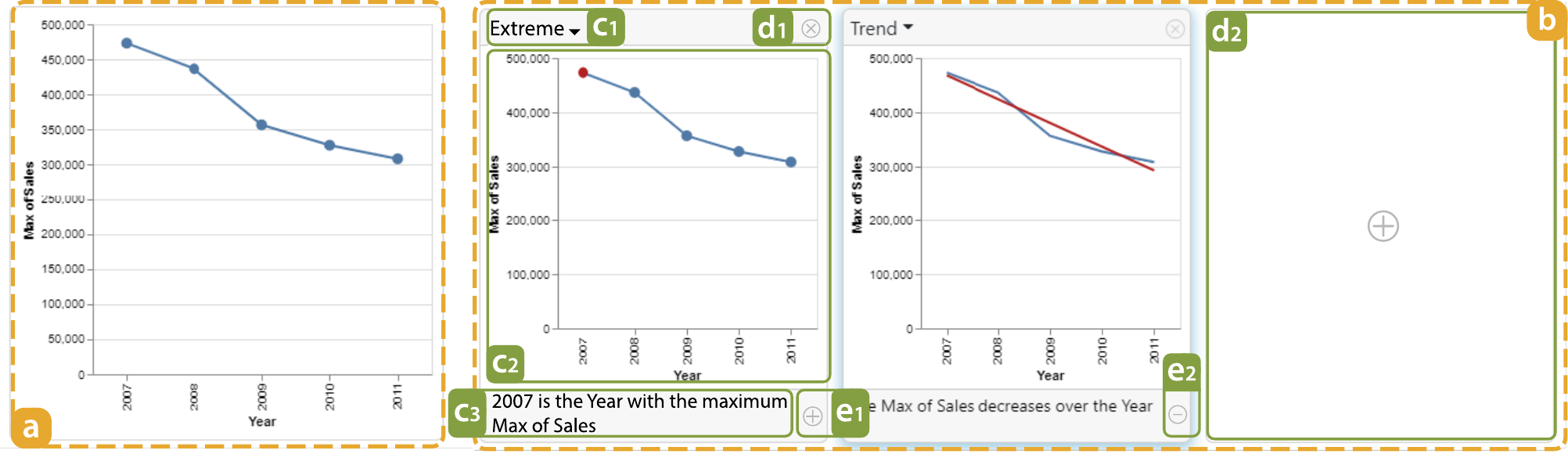}
    \caption{This figure demonstrates a plot widget in \tool. It contains the original plot in (a) and suggested data facts in (b).
    The plot widget supports fact customization with (c1)-(c3).
    (d1) and (d2) allow users to delete existing facts and create new facts.
    (e1) and (e2) enable adding and removing facts from the story.
    }
    \label{fig:plot_widget}
     \Description{This figure demonstrates a plot widget in Notable. It contains the original plot in (a) at the leftmost position in the plot widget. The illustrated facts are placed at the right in (b) and can be customized with interactions. In each fact, a user can change the fact type through a dropdown list in (c1) at the top, highlight data points by clicking on the charts in (c2), and revise the text description using the text entry box in (c3) at the bottom. The user can also remove an existing fact in (d1) at the top right corner or add a customized fact by creating a new one in an empty slot in (d2) on the right. The user can click the buttons (e1) and (e2) at the bottom right corner to add a fact to the story or remove it from the story.}
\end{figure*}

\subsubsection{Plot Widget}
As shown in Figure~\ref{fig:interface}, one plot widget appears after the input cell. 
It is designed to support browsing, editing, and selecting facts for storytelling during exploration~(\textbf{R1}).
A plot widget presents the original chart at the leftmost position~(Figure~\ref{fig:plot_widget}(a)) and illustrated data facts at the right~(Figure~\ref{fig:plot_widget}(b)).
The data facts follow a decreasing order of their scores.
The layout is inspired by Lux~\cite{lee2021lux}.
Users can scan through all suggested data facts without additional interaction.
The card of a data fact contains the fact type at the top~(Figure~\ref{fig:plot_widget}(c1)), the illustrated chart in the middle~(Figure~\ref{fig:plot_widget}(c2)), and the text description at the bottom~(Figure~\ref{fig:plot_widget}(c3)). 
\revision{Users are also allowed to select
the fact type with a dropdown list, modify the text description in a text entry box, and click on the data point to highlight it. Such interactions enable users to create customized facts by assigning fact types, documenting related findings, and highlighting key data points. These customized facts can be added to the story.}
\tool currently supports manually highlighting a single data point and can be extended to multi-point selection in the future.
Users can click the cross icon~\includegraphics[height=0.8em]{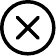} at the top right corner~(Figure~\ref{fig:plot_widget}(d1)) to delete a fact and click on the plus icon~\includegraphics[height=0.8em]{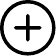} in the middle of an empty card~(Figure~\ref{fig:plot_widget}(d2)) to add a fact.
When a useful data fact is noticed, clicking the plus icon~\includegraphics[height=0.8em]{figure/icon/plus-circle.pdf} at the bottom right corner can add the fact to the story~(Figure~\ref{fig:plot_widget}(e1)).
Once a fact is in the story, its card will be highlighted with a blue shadow, and the plus icon~\includegraphics[height=0.8em]{figure/icon/plus-circle.pdf} turns into a minus icon~\includegraphics[height=0.8em]{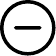}~(Figure~\ref{fig:plot_widget}(e2)).
Clicking the minus icon will remove the fact from the story.

\subsubsection{Organization Panel}
\begin{figure}[h]
    \centering
    \includegraphics[width=\linewidth]{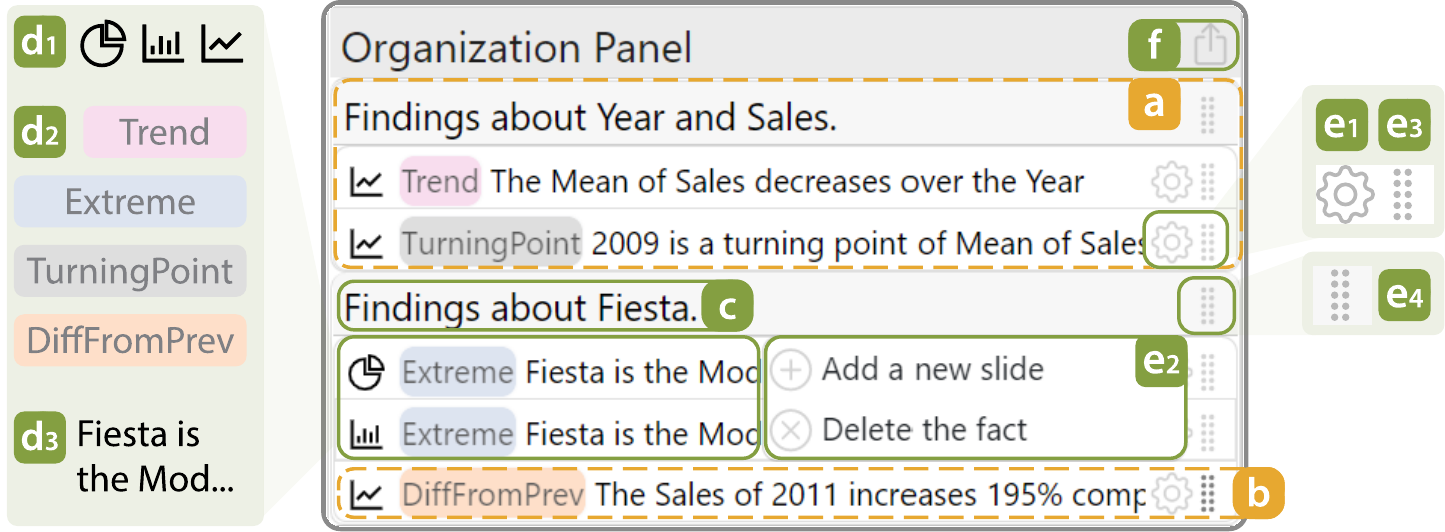}
    \caption{This figure demonstrates the organization panel in \tool. \revision{In this figure, the center part is the organization panel in \tool. The left and right parts with a green background are for illustration.} The organization panel provides an outline view of the organized story and supports story adjustment. (a) and (b) represent a slide and a fact, respectively. (c) is the slide title. (d1)-(d3) show the chart type, fact type, and description of a fact. (e1)-(e4) enable fact sequence arrangement. (f) is for exporting the slides.
    }
    \Description{This figure demonstrates the organization panel in Notable. In this figure, the center part is the organization panel in Notable. The left and right parts with a green background are for illustration. The organization panel provides an outline view of the organized story and supports story adjustment. (a) at the top and (b) at the bottom represent a slide and a fact, respectively. (c) in the middle is the slide title. (d1)-(d3) on the right show the chart type, fact type, and description of a fact. (e1)-(e4) on the left enable fact sequence arrangement. (f) at the top right corner is for exporting the slides.}
    \label{fig:organization_panel}
\end{figure}

When data facts are selected, they will be added to the story, and the organized story is shown in the organization panel.
The organization panel presents the outline of the story and allows customization~(\textbf{R4}), as shown in~Figure~\ref{fig:organization_panel}.
It also supports the need for convenient fact documentation~(\textbf{R2}) by presenting users' selected facts.
Each card in the organization panel represents a slide~(Figure~\ref{fig:organization_panel}(a)), while each list item in a card encodes a fact~(Figure~\ref{fig:organization_panel}(b)).
The top part of each card is a modifiable slide title~(Figure~\ref{fig:organization_panel}(c)).
Due to screen space constraints, the illustrated data fact can hardly be presented in the organization panel.
To mitigate the issue, a glyph of chart type~(Figure~\ref{fig:organization_panel}(d1)), the fact type~(Figure~\ref{fig:organization_panel}(d2)), and the text description~(Figure~\ref{fig:organization_panel}(d3)) are provided.
Similar to plot widgets, the fact description can also be customized.
Furthermore, when clicking the gear icon~\includegraphics[height=0.8em]{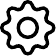}~(Figure~\ref{fig:organization_panel}(e1)), more operations are provided, \ie, removing the fact from the story and creating a new slide for the fact~(Figure~\ref{fig:organization_panel}(e2)).
The organization panel supports the sequence adjustment by dragging facts or slides and dropping them to the desired positions with the grip icons~(Figures~\ref{fig:organization_panel}(e3) and (e4)).
When users are satisfied with their slides, they can export the slides by simply clicking the icon~\includegraphics[height=0.8em]{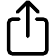} at the top right corner of the panel~(Figure~\ref{fig:organization_panel}(f)).

When designing the organization panel, we first thought about showing the story with a tree of facts~\cite{obie2022gravity++, mathisen2019insideinsights}.
However, one concern of tree-based design is the scalability issue~\cite{heer2004doitrees, tsang2020tradao}.
When the story has many facts, it will be challenging to display them.
Furthermore, since the slideshow follows a linear sequence, the users need additional mental effort to map a hierarchical tree of facts to a linear presentation.
\revision{Another possibility is to show all output slides~\cite{zheng2022telling} and allows direct slide manipulation.
It facilitates an intuitive preview of generated slides.
However, given the limited screen space, the users can hardly gain an overview of the story and adjust the sequence.
Furthermore, \tool, as a lightweight extension, is challenging to provide the complete functionality of slide editing as PowerPoint does.
Therefore, we decided to present the story's overview, similar to PowerPoint's outline view.
Such a design shows the complete sequence of facts while preserving the hierarchical relationship between facts and slides.
At the same time, some basic functionalities, such as revising the description, are provided.
Users can improve the slides further, such as adding animations, with their familiar presentation tools~(\textbf{R5}).
In the future, it might be more helpful to provide both the outline view and the slide view in \tool and support more functionalities.}

\section{Usage Scenario}

In detail, we describe a usage scenario to illustrate how \tool assists data storytelling in computational notebooks.
Imagine Dora, a business analyst from BMW who plans to explore the car sale dataset and report findings in an upcoming meeting.
The dataset records the sales of cars from different brands and categories within five years.
There are five columns, including one quantitative attribute (\textit{Sales}), three categorical attributes (\textit{Brand}, \textit{Model}, \textit{Category}), and a temporal attribute (\textit{Year}).

\begin{figure*}
    \centering
    \includegraphics[width=0.9\linewidth]{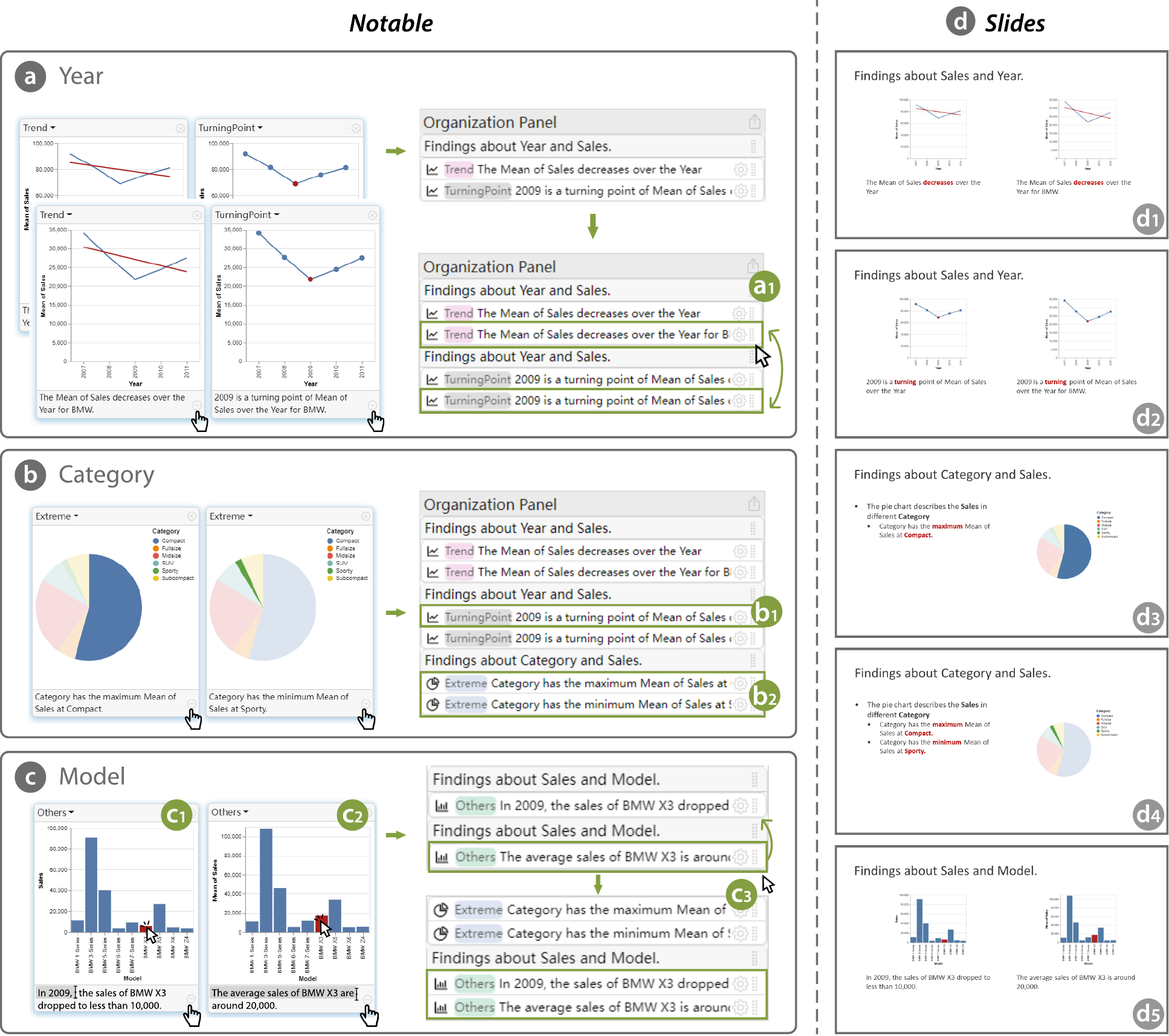}
    \caption{This figure illustrates the usage scenario. (a)-(c) demonstrate the procedure of creating a data story with \tool. (d) shows the exported slides. 
    \revision{(d1) and (d2) correspond to the facts identified in (a); (d3) and (d4) are generated based on (b); and (d5) shows the facts in (c).}} 
    \Description{This figure illustrates the usage scenario. (a)-(c) on the left demonstrate the procedure of creating a data story with Notable. (d) on the right shows the exported slides. From the top to the bottom of (d), (d1) and (d2) correspond to the facts identified in (a); (d3) and (d4) are generated based on (b); and (d5) shows the facts in (c).}
    \label{fig:scenario}
\end{figure*}

\textbf{Exploring data by creating charts. }
To gain an overview of BMW sales, Dora first probes into the relationship between time and sales with a line chart (Figure \ref{fig:scenario}(a)).
Based on the chart, \tool illustrates three data facts regarding the difference between two consecutive years, the overall trend, and the turning point.
Dora quickly scans through the illustrated facts and selects the fact regarding the trend to give a big picture of sales.
She also notices the year 2009 as a turning point over the five years, which may be worth reporting.
She adds the fact into her story.
Dora plots the sales of all brands over the past years to learn whether the year 2009 is only a turning point of BMW sales or all brands’ sales. 
She finds actually all brands' sales experience a similar trend and has the year 2009 as a turning point.
To facilitate the comparison, she moves the facts about data trends and turning points and places them with corresponding BMW's facts in the same slide~(Figure~\ref{fig:scenario}(a1)).
Next, Dora investigates the sales in 2009 as it is a turning point.
She checks the relationship between categories and sales (Figure~\ref{fig:scenario}(b)).
\tool highlights two facts about extreme values, the maximum at compact models and minimal at sporty models.
By comparing the two facts, she realizes the unbalanced sales of different categories.
Both facts are included in her story.
She also wonders about the sales of car models in 2009.
She plots a bar chart and notices BMW Z4 is the model with the worst sales (Figure~\ref{fig:scenario}(c1)).
The fact about Z4 is added to her story as well.
Then Dora reads her story in the organization panel.
The story starts with an overview and then drills down to the sales by categories and models in 2009, a turning point.
A question comes to her: \textit{``do BMW Z4's low sales lead to the overall unsatisfactory sales in 2009?''}
She plots the relationship between models and average sales over the past years.
However, there is no obvious difference between BMW Z4's sales in 2009 and its average sales (Figure~\ref{fig:scenario}(c2)).
Therefore, her hypothesis is rejected.
Furthermore, she notices BMW X3 has unusually low performance in 2009 compared to its average sales.
The finding may help explain the overall low sales in 2009.
To report the finding, she creates two new facts about BMW X3 with \tool and writes down her observation.
The bars of BMW X3 are highlighted by clicking on them (Figure~\ref{fig:scenario}(c1) and (c2)).
Finally, she adds both facts to her story and ends her exploration.

\textbf{Organizing data facts and exporting. }
While adding data facts into the organization panel, Dora observes that \tool organizes facts automatically. 
She notices that the sequence of slides follows the drill-down pattern of data stories~(Figure~\ref{fig:scenario}(b2)).
The slide about sales and categories in 2009 is after the slides that introduce the trend over five years.
Moreover, Dora notices that \tool generates slide titles (e.g., ``Findings about Sales and Year'') simultaneously  (Figure~\ref{fig:scenario}(b1)).
After exploration, satisfied with most slide sequences, 
Dora merges two facts about BMW X3 in one slide (Figure~\ref{fig:scenario}(c3)) and removes the fact about BMW Z4.
Ultimately, Dora clicks the export button and downloads the slides (Figure~\ref{fig:scenario}(d)).
She adjusts the style of slides with PowerPoint and shares them with her team.
\section{User Study}
We conducted a user study
to verify the effectiveness and usability of \tool.
The setup of our user study is introduced in Sections~\ref{sec:user_participant}-\ref{sec:user_procedure} and the results are reported in Section~\ref{sec:result}.

\begin{figure*}
    \centering
    \includegraphics[width=0.9\linewidth]{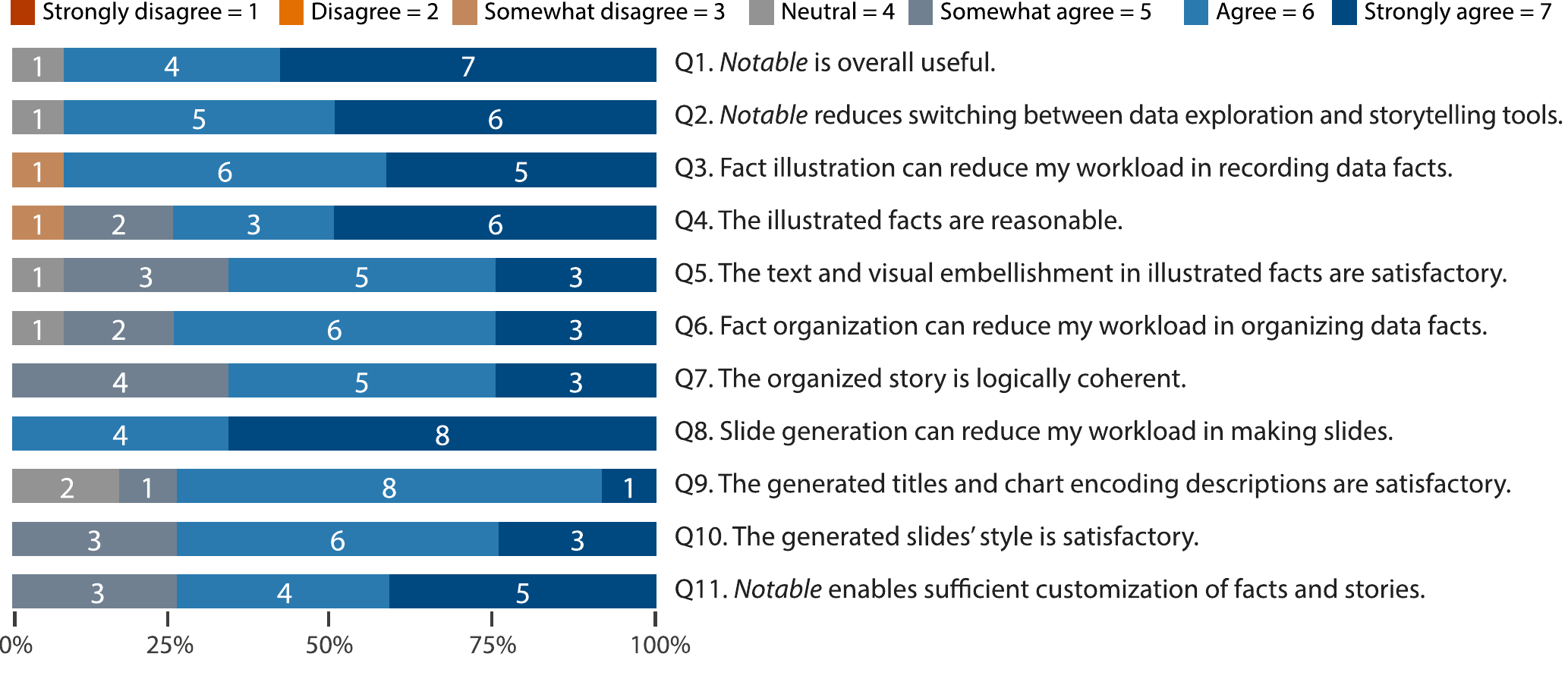}
     \caption{This figure shows the results of our quantitative evaluation of the effectiveness. 
     }
    \label{fig:quantitative_result}
    \Description{This figure shows the results of our quantitative evaluation of the effectiveness. A stacked bar chart shows the distribution of scores on the left. The corresponding question is placed next to each bar.}
\end{figure*}

\subsection{Participants}\label{sec:user_participant}
In our user study, we recruited 12 data workers (11 male and 1 female, $Age_{mean} = 28.75$, $Age_{std} = 5.90$, $Experience_{mean} = 5.67~years$, $Experience_{std} = 2.72~years$) from our institution through social media and word-of-mouth (denoted as U1-U12).
They were one software engineer~(U1), five postgraduate researchers~(U2, U3, U6, U7, U9), three applied data scientists~(U4, U10, U12), two research scientists~(U5, U11), and one product manager~(U8).
Their experiences in data analysis ranged from two years to ten years.
Since \tool is developed based on Python and JupyterLab, we required the participants to be familiar with them as well.

\subsection{Task and Dataset}\label{sec:user_tasks}
In our user study, we designed an open-ended data analysis task that involves data exploration and storytelling.
Participants were asked to explore a given dataset with charts and prepare a slide deck to tell a data story about the findings.
We required that the participants explored the dataset with at least six charts and made a story with at least five charts.
By doing so, we would like to ensure that participants explored the dataset with \tool sufficiently and were able to make a story.
We selected \textit{movies} dataset from Vega Datasets\footnote{\url{https://github.com/vega/vega-datasets/}}, which have been widely used in evaluating the effectiveness and usability of intelligent visualization tools, including Voyager2~\cite{wongsuphasawat2017voyager2} and DashBot~\cite{deng2022dashbot}.
Another advantage of using \textit{movies} dataset is that its data attributes are familiar to the general public~\cite{saket2018task}.
Therefore, the performance of users will not be affected by their limited knowledge of datasets.  
To control the time of our user study, we reduced the dataset size to 9 attributes and 392 rows.
The remaining attributes still covered three major types of tabular data~(\ie, nominal, quantitative, and temporal).
We also removed null values to eliminate the workload of data cleaning and let the participants focus on data exploration and storytelling.

\subsection{Procedure}\label{sec:user_procedure}
All studies were conducted through one-on-one offline meetings.
Before the user study, we briefed the procedure and asked for the participants' consent to record the study.
In the user study, we first introduced different components of \tool and related interactions.
Then the participants were asked to try \tool for around 10 minutes until they felt familiar with it.
Next, we asked the participants to finish the task using \tool.
The participants ended the task until they were satisfied with their slides.
The task took around 30 to 40 minutes. 
In the end, we interviewed the participants regarding their experiences.
They were also asked to fill a 7-point Likert questionnaire to rate their effectiveness and usability.
In the questionnaire, 1 point means ``strongly disagree'' while 7 points mean ``strongly agree''.
The eleven questions about the effectiveness are shown with results in Figure~\ref{fig:quantitative_result}.
Q1 and Q2 evaluate the overall performance of \tool.
Q3-Q5, Q6-Q7, and Q8-Q10 evaluate the fact illustration, fact organization, and slide generation modules, respectively.
Q11 asks whether \tool enables users' sufficient customization.
The questions regarding usability were from System Usability Scale~(SUS)~\cite{brooke1996sus}, a widely adopted approach to measure the usability of an application.
It took around 20 to 30 minutes to finish the interview and the questionnaire.
The whole user study lasted about 1-1.5 hours.
Each participant received \$7.5 as compensation.
The authors took notes to record feedback during the study.

\subsection{Results}\label{sec:result}
In this section, we report the quantitative and qualitative results (Sections~\ref{sec:user_quan_result} and~\ref{sec:user_qual_result})  of the user study.

\subsubsection{Quantitative results.}\label{sec:user_quan_result}
The quantitative results of our user study reflect participants' ratings on both effectiveness and usability.
The effectiveness ratings are shown in Figure~\ref{fig:quantitative_result}.
As the results indicate, most of the participants felt satisfied with \tool. 
They agreed that the three key computation modules, fact illustration, fact organization, and slide generation, were useful and able to achieve their expectation.
Furthermore, we noticed that all participants agreed that \tool had sufficient support for story customization~(\textbf{R4}).
The usability score of \tool reaches 86.1, which is higher than 95\% of applications, according to Sauro and Lewis~\cite{sauro2012quantifying}.
The results show that the participants thought highly of the usability and, the interactive modules were intuitive to them.

\subsubsection{Qualitative results.}\label{sec:user_qual_result}
This section 
reports participants' qualitative feedback about the overall experience, illustrated facts, organized stories, and generated slides.
To derive the qualitative results, two co-authors summarized the participants' feedback after reading the recording transcripts and the notes taken during the study individually.
Then co-authors discussed together and reached a consensus on the common findings revealed in the user study.

\revision{\textbf{\tool~can help users create data stories during data exploration.}}
All participants appreciated the overall experience of using \tool. 
For example, U1 expressed his feeling by saying that \textit{``the tool is amazing''}. 
U2 could not wait for the release of \tool and said that he would like to install it soon.
They believed that the workflow of \tool is reasonable and 
\revision{has the potential to} reduce their workload in creating data stories with findings from data exploration.
For example, U12 said \textit{``the tool can integrate data processing, data exploration, chart plotting, and slide generation and thus reduces switching between multiple tools''}.
U10 expressed a similar opinion 
by saying \textit{``the overall workflow is promising''}.
Their opinion verified that the design requirement \textbf{R1} had been fulfilled.
Furthermore, they felt that \tool is integrated into computational notebooks seamlessly.
U12, as an experienced commercial software developer, believed its good integration into a commonly used notebook environment could attract a broad range of users.

\textbf{Illustrated facts \revision{have the potential to} benefit users from multiple perspectives.}
Among all functionalities in \tool, fact illustration has been mentioned frequently as one of \tool's most helpful functions.
Most of the participants agreed that automatically illustrated facts are satisfactory and 
\revision{are likely to} reduce their efforts in recording and highlighting data findings~(\textbf{R2}).

Besides the reduction of fact recording efforts, some other benefits brought by fact illustration were also noticed.
First, U1, U2, and U8 commented that the illustrated facts revealed widely examined data patterns such that they could save time in writing and running the analysis code.
For example, when data trend is illustrated in the charts, the regression analysis may be eliminated.
U8 said \textit{``the automatic extreme fact generated by the system is a great help, especially for huge datasets''}.

\begin{figure*}
    \centering
    \includegraphics[width=\linewidth]{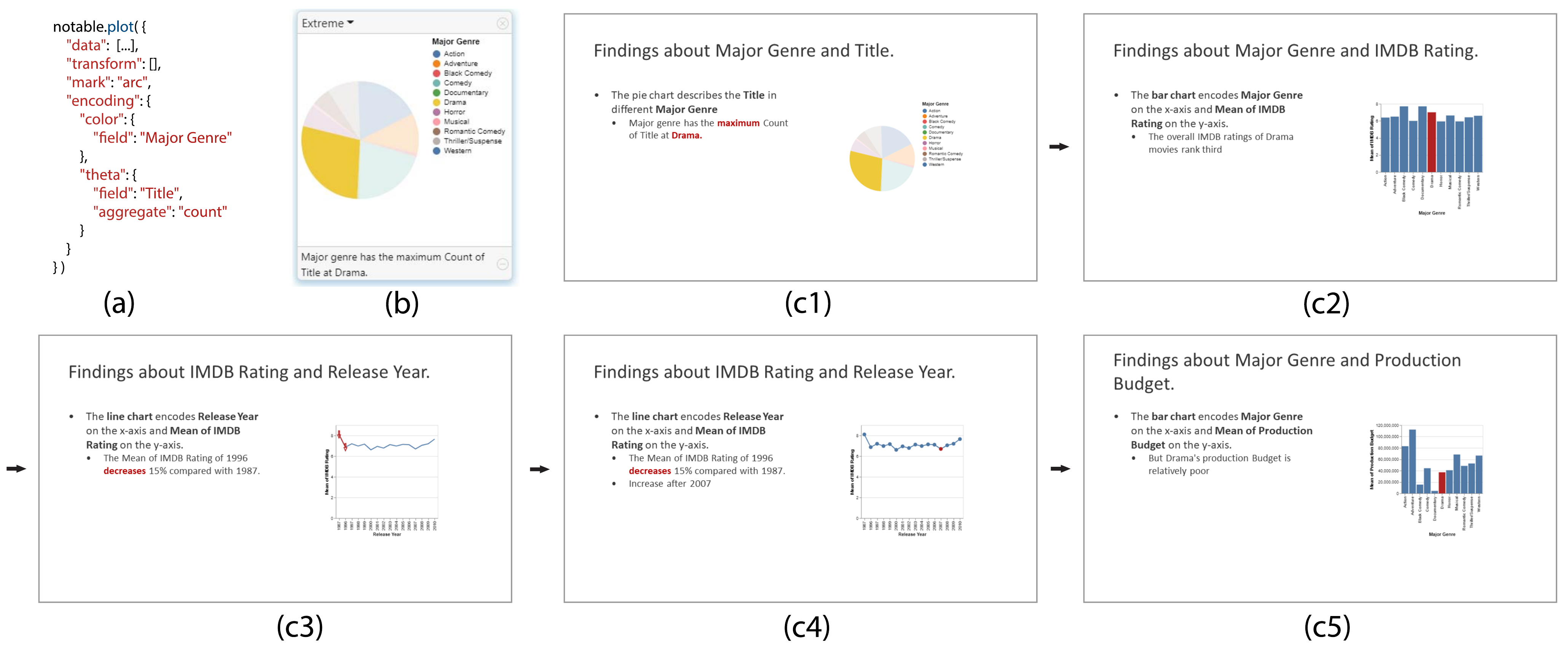}
    \caption{This figure presents a case in our user study. When the participant used (a) as inputs, \tool illustrated a data fact (b). In the fact, drama was highlighted as the genre which has the most movies. The data fact guided the participant to conduct a series of explorations on drama movies. The participant finally created a data story as (c1)-(c5) show.}
    \label{fig:case1}
    \Description{This figure presents a case in our user study. When the participant used (a) at the top left corner as inputs, Notable illustrated a data fact (b) next to (a). In the fact, drama was highlighted as the genre which has the most movies. The data fact guided the participant to conduct a series of explorations on drama movies. The participant finally created a data story as (c1)-(c5) from the top middle part to the bottom right part show.}
\end{figure*}

Second, U6 and U11 also mentioned that the illustrated data facts could guide his data exploration.
When U6 started exploring the movie dataset in our user study, he did not have a clear idea about what interesting story might be distilled from the data.
The illustrated data facts (Figure~\ref{fig:case1}(b)) served as \textit{``hints''} to him and guided him to continue the exploration of movies that belongs to the genre of drama.
Finally, he was able to create a complete data story as Figures~\ref{fig:case1}(c1)-(c5) show.
In the story, he first identified drama as the most frequent genre~(Figure~\ref{fig:case1}(c1)).
Then he reported that the average rating of drama movies is the third highest among all genres~(Figure~\ref{fig:case1}(c2)).
The trend of ratings of drama movies was further shown in the next two slides~(Figures~\ref{fig:case1}(c3) and (c4)).
At the end of the slide deck, he highlighted that the production cost of drama films was low though their ratings were great, to attract the audience's interest~(Figure~\ref{fig:case1}(c5)). 

Though the fact illustration function was 
\revision{thought highly}, it has room for future improvements.
First, the diversity and complexity of illustrated facts should be enhanced in the future.
U10 was the only user who expressed unsatisfactory with the performance of fact illustration in Q3 and Q4 of the questionnaire~(Figure~\ref{fig:quantitative_result}).
The reason was that he felt the facts were not \textit{``in-depth''}. 
In the user study, he attempted to identify whether the gross of a film was predictable and then presented the results to his team.
First, he identified that the relationship between movie genres and average gross might be used for prediction.
Then he further confirmed whether the distribution of movie gross was concentrated in each genre and whether the sample size of each genre was large enough.
Such results could help him to determine whether the relationship was robust.
However, \tool failed to consider these analyses, and thus he had to record data findings manually.
He suggested that \tool can be further improved to consider the complex and diverse insights according to the users' intention, such as identifying relationships for prediction.
U3 and U5 also expressed their expectation for more diverse facts such as data clusters.
Second, the description can be improved.
Though most of the users were satisfied with the results (Figure~\ref{fig:quantitative_result}), U2, U6, and U7, as experts in natural language processing, pointed out that our template-based descriptions were less flexible and thus resulted in some unnatural expressions, \eg, ``Director has the maximum Mean of US Gross at James Cameron''.
To address the issue, they suggested the usage of advanced language generation models, such as T5~\cite{raffel2020T5}, but also warned that such models were less controllable and might increase the latency of description generation.

\textbf{Fact organization facilitates more than storytelling.}
\revision{According to our observation, the participants followed the suggested data fact organization most of the time and did not often arrange facts in stories manually.
The observation is also supported by the quantitative results in Figure~\ref{fig:quantitative_result}.}
The results show that our organized stories were generally considered logically coherent~(\textbf{R3}) and users' effort in creating stories \revision{is possible to be} eliminated.

On top of its advantages to storytelling, we received feedback in which the fact organization function and the organization panel were reported to benefit their data exploration.
For example, U7 felt that the panel could give him an overview. 
U4 commented that \textit{``the organization panel helps taking notes and collecting interesting insights''} and \textit{``(the organized story) facilitates and guides the data exploration in the next step''}.
U6's comments supplemented U4's comment by mentioning that the organized story helped him identify logical flaws in the exploration.
U1 and U10 further emphasized the importance of presenting organized facts in the organization panel when they conducted complex data analysis.
U10 mentioned that \textit{``the panel will be really helpful when conducting complex data analysis, especially those analyses where I need to explore data back and forth. For example, it helps track what directions I have explored''}.
Such comments verify that organizing data facts into stories can have a positive effect on data exploration as well, which demonstrates the value of providing on-the-fly assistance to storytelling during data exploration (\textbf{R1}).

We also notice one potential improvement on \tool's fact organization module.
As introduced in Section~\ref{sec:fact_organization}, our algorithm of fact organization minimizes the transition cost of data stories.
However, the semantic relationship between data facts is not thoroughly considered and thus leads to a suboptimal case.
In U6's data story (Figure~\ref{fig:case1}), he would like to express that the budgets of drama movies were low though the ratings were high using slides in Figures~\ref{fig:case1}(c2)-(c5).
However, the fact organization module did not consider the text description and placed the slide in Figure~\ref{fig:case1}(c5) between slides in Figures~\ref{fig:case1}(c1) and (c2).
The reason was that slides in Figures~\ref{fig:case1}(c1), (c2), and (c5) were about movies of all genres, while slides in Figures~\ref{fig:case1}(c3) and (c4) only concerned drama movies.
The fact organization module considered the analysis from slides in Figures~\ref{fig:case1}(c1), (c2), and (c5) to slides in Figures~\ref{fig:case1}(c3) and (c4) as a drill-down analysis and therefore arranged the sequence in a sub-optimal approach.
U6 had to arrange the facts according to the semantic meanings manually.
Such a mistake reveals the necessity of considering user-specified semantic information of data facts in future improvements.

\textbf{Generated slides will be better with a more personalized design.}
The slide generation function was appreciated by the participants since it 
\revision{is likely to reduce} their effort in making neat slides with data findings.
U11 described the functionalities as \textit{``one-click generation''}, which demonstrates that he considered our slide generation as a 
convenient function.

However, regarding the style of slides, different participants hold diverse opinions.
For example, U2 appreciated the generated slides since they were \textit{``organized and neat''}.
The highlighted texts in the slides were well received by U9.
However, U1 expressed a different opinion.
Though he agreed that the generated slides can be presented in informal meetings, 
he would spend great effort to beautify them before presenting them to his supervisor.
Furthermore, different users had different opinions about the quantity of information in generated slides.
For example, U10 considered the chart description is too long. 
In practice, he only wrote bullet points such as \textit{``column A vs. column B''}, \eg, \textit{``Major Genre vs. US Gross''}.
On the contrary, U5 commented that the current design is less informative than his slides.
He preferred to introduce more about data facts with images and texts.
According to the feedback, we summarize two implications for future tools.
First, to fulfill various requirements, the tool should provide diverse slide templates including the content and style.
For example, chart descriptions can be bullet point-style or sentence-style.
Second, it is important to allow future modification of exported slides with presentation tools.
Even though diverse templates are provided, it may not be possible to fulfill users' personal needs.
For example, U9 mentioned that he would like to add animation to slides.
Therefore, facilitating further improvements is necessary.
\section{Discussion}
In this section, we first highlight two lessons learned from the research and future directions about \textit{human-machine collaboration in \tool}~(Section~\ref{sec:lesson_hai_collaboration}) and \textit{connection between data exploration and storytelling }~(Section~\ref{sec:lesson_exploration_story}).
Then the limitation of our research is discussed~(Section~\ref{sec:limitation}).

\subsection{Human-machine collaboration in \tool}\label{sec:lesson_hai_collaboration}
\revision{In our paper, we propose \tool, a computational notebook extension, to introduce machines to the loop of data exploration and storytelling.
According to our evaluation results, the workflow involving \tool is highly appreciated by participants since it both provides necessary assistance, such as highlighting data points and transferring data, and allows humans to control the entire process.}
With \tool, the user is responsible for deciding what data to be explored while \tool only illustrates the potentially interesting data facts according to the user-created charts (Section~\ref{sec:fact_illustration}) and organizes user-selected data facts.
After that, the user takes charge of reviewing the logical flow of the story when creating stories.
On the other side, machines take responsibility for several repetitive tasks, such as transferring data from the exploration stage to the storytelling stage and highlighting key data points. Though these tasks seem not to be challenging, they actually take great effort from users, according to our formative interviews~(Section~\ref{sec:formative}). Furthermore, potential errors that may be introduced by humans in data transferring could be avoided~\cite{brehmer2021jam}.

We also receive some comments on refining the workflow.
U4 proposed that the machine could take some more steps to facilitate users further.
He thought that \tool should suggest more related data facts when a data fact is selected, similar to the recommendation in Lux~\cite{lee2021lux}
\revision{and Erato~\cite{sun2022erato}}.
He also mentioned that \tool might recommend and change the chart types of the original chart to present data facts more effectively.
If these changes are applied, we consider that machines are not only responsible for repetitive tasks but also attempt to guide or correct humans, which can be ``double-edged swords''~\cite{li2021exploring}.
When machines take further steps, humans need to spend extra mental load to understand why these steps were taken. When the results are worse than expected, humans may lose confidence in machines. The bias led by machines should also be aware.
In the future, we will continue the research on human-machine collaboration to study how to maximize the values of machines and humans in data storytelling.

\subsection{Connection between data exploration and storytelling}\label{sec:lesson_exploration_story}
Data exploration and storytelling are two necessary stages that are closely connected theoretically. 
However, previous studies have mentioned that they are loosely connected practically, considering the gap in converting analysis results to data stories~\cite{gratzl2016visual, brehmer2021jam, chevalier2018analysis}. 

From the feedback in the evaluation, we find \tool \revision{ has the potential to} help users quickly create a data story in the form of presentation slides simply with several clicks (Section~\ref{sec:result}), which matches the design requirements of \tool.
On top of that, it is interesting to verify that presenting the organized story in the organization panel has a positive influence on data exploration.
Several user study participants indicated that the organized story could help them identify the logical flaw and remind them of explored facts~
(Section~\ref{sec:user_qual_result}).
Feedback from our user study indicates that the design of \tool not only facilitates the connection between the two stages but further enhances the bi-directional connection of both data exploration and storytelling. 
It also informs us of the necessity of enhancing the bond between them in future tools.
In the future, some steps can be taken toward bridging the gap between exploration and storytelling.
In the formative study, P4 pointed out that he would like to transfer data findings among various data exploration tools~(\eg, Stata\footnote{\url{https://www.stata.com/}}).
Such requirements encourage us to develop multiple versions of \tool for different data exploration tools. 
Our computation modules can be re-used and the interactive modules require refinement to fit different interfaces.
Furthermore, the support of other storytelling formats, such as reports, is possible to be integrated into \tool by augmenting the slide generation module.
It will also be interesting to investigate the factors that affect the bi-directional connection between exploration and storytelling such as the approach of presenting data stories in storytelling tools.
The results can be applied to optimize the design of \tool.

\subsection{Limitations}\label{sec:limitation}
Our research is not without limitations. In this section, we discuss the limitations of our research from the functionalities and the evaluations of \tool.

\subsubsection{Functionalities}
The functionalities of \tool can be further extended.
First, \tool is limited by supporting data facts in basic charts.
Currently, \tool only supports five basic chart types: bar chart, pie chart, line chart, area chart, and scatter plot.
As mentioned by user study participants, U3 and U10, the support to other chart types (\eg, heatmaps and box plots) and multi-view visualizations could be introduced.
\revision{Second, the consideration of user-customized facts is limited.
As mentioned in Section~\ref{sec:interface}, users are allowed to modify the illustrated facts and create new facts.
However, \tool is not able to understand users' input completely.
It cannot infer three attributes in a data fact, parameters, the focus, and the score.
Therefore, the organization of these customized facts will only consider the other four attributes.
Furthermore, the semantic information of created descriptions is not considered in fact organization.
As described in Section~\ref{sec:user_qual_result}, failing to consider the semantic information might lead to some suboptimal fact sequences. 
In the future, we plan to extend \tool further to handle users' input more comprehensively and improve its functionalities.}

\subsubsection{Evaluations}
\revision{We conducted an in-lab user study where the participants explored the \textit{movies} dataset and created a data story with \tool.
There are three perspectives to improve the evaluation.
First, comparing \tool with users' familiar real-world workflow may reveal more insights, such as differences in slide quality, preparation time, and the workload of creating slides.
Second, more long-term evaluation is desired.
The \textit{movies} dataset may not be as complex as real-world datasets.
The participants commonly finished the task in 30 to 40 minutes.
It will be interesting to learn users' feedback when \tool is applied in their daily work for a longer period.}
Finally, the participants in both the formative study and the user study have limited coverage.
Though we have attempted to improve the diversity of participants, \eg, by recruiting participants with diverse backgrounds and various experiences, we acknowledge some limitations, including the imbalance of gender distribution in the user study and the missing of some types of data workers in the formative study~(\eg, data journalists).
In the future, we hope to deepen our understanding of data exploration and storytelling in the long-term real-world usage of \tool by diverse users.
\section{Conclusion}
To communicate data findings in computational notebooks, users have to spend considerable effort in turning them into data stories.
In our research, we explore offering on-the-fly assistance to users to facilitate effective data storytelling during data exploration.
We first conducted formative interviews with data analysts with diverse backgrounds to derive the design requirements.
Then based on the requirements, \tool, a computational notebook extension, is proposed to facilitate fact documentation and organization with intelligent support.
\tool was generally appreciated by the users in a user study with 12 data workers.
In the future, we hope to further improve \tool by considering more data fact types (\eg, data clusters) and enabling personalized slide generation.
\revision{It will also be interesting to investigate other approaches to reduce users' burden, such as simplifying the input format and recommending facts based on users' preferences.}
Furthermore, a long-term evaluation has the potential to reveal its pros and cons in a real-world setting.

\begin{acks}
The authors would like to thank the reviewers, Aoyu Wu, and Liwenhan Xie for their constructive suggestions and all participants in our studies.
The research was partially supported by the Hong Kong Research Grants
Council (GRF16210722) and the National Natural Science Foundation of China (U22A2032).

\end{acks}

\balance

\bibliographystyle{ACM-Reference-Format}
\bibliography{main}

\end{document}